\newif\ifdraft
\draftfalse

\newif\iftwocolumn
\twocolumnfalse

\ifdraft
  \documentclass[12pt, twocolumn]{article}
\else
  \documentclass[twocolumn]{article}
\fi

\usepackage[margin=1in]{geometry}
\usepackage[parfill]{parskip}
\usepackage[utf8]{inputenc}
\usepackage{hyperref}
\usepackage{titlesec}

\iftwocolumn
\else
\fi

\titleformat{\subsection}[runin]{\bfseries}{\thesubsection}{1em}{}
\renewcommand{\thesubsection}{%
  \arabic{section}.\arabic{subsection}.{}}

\usepackage[square,numbers]{natbib}

\usepackage{amsmath,amssymb,amsfonts,amsthm}

\usepackage{graphicx}

\usepackage[export]{adjustbox}
\graphicspath{ {images/} }
\let\oldincludegraphics\includegraphics
\renewcommand{\includegraphics}[2][]{%
  \oldincludegraphics[#1,max width=\linewidth]{#2}}

\usepackage{abstract}

\renewenvironment{abstract}{%
  \list{}{%
    \iftwocolumn
        \setlength{\leftmargin}{8em}
    \else
        \setlength{\leftmargin}{2em}
    \fi
    \setlength{\rightmargin}{\leftmargin}
  }
  \item\relax
}
{\endlist}

\ifdraft
  \usepackage{lineno}
  \usepackage{xcolor}
  
  \linenumbers
\fi

\title{\textbf{(Un)paired signal-to-signal translation with 1D conditional GANs}}

\author{Eric Easthope \\
        \small{University of British Columbia} \\
        \small{Department of Electrical and Computer Engineering} \\
        \small{Vancouver, British Columbia, Canada}
        }

\date{}

\begin{document}

\onecolumn
\maketitle

\thispagestyle{empty}

\begin{abstract}

  
  I show that a one-dimensional (1D) conditional generative adversarial network (cGAN) with an adversarial training architecture is capable of unpaired signal-to-signal (\texttt{sig2sig}) translation. Using a simplified CycleGAN model with 1D layers and wider convolutional kernels, mirroring WaveGAN to reframe two-dimensional (2D) image generation as 1D audio generation, I show that recasting the 2D image-to-image translation task to a 1D signal-to-signal translation task with deep convolutional GANs is possible without substantial modification to the conventional U-Net model and adversarial architecture developed as CycleGAN. With this I show for a small tunable dataset that noisy test signals unseen by the 1D CycleGAN model and without paired training transform from the source domain to signals similar to paired test signals in the translated domain, especially in terms of frequency, and I quantify these differences in terms of correlation and error.
\end{abstract}

\clearpage
\pagenumbering{arabic}

\iftwocolumn
  \twocolumn
\else
  \onecolumn
\fi

\section{Introduction}
\subsection{Background.}
The past few years have seen a significant rise in research and public interest in the use of generative machine learning and artificial intelligence (ML/AI) models for image-to-image translation tasks.

Perhaps one of the more recognizable models is \texttt{pix2pix} \cite{pix2pix}, a deep generative model (DGM) and particularly a deep convolutional generative adversarial network (DCGAN) \cite{GAN, DCGAN} that is capable of translating between pairs of high-resolution images within a learned image data domain. The novelty of pix2pix laid in its model architecture which combined a deep U-Net generator that learns to generate mock data samples with a convolutional PatchGAN discriminator that learns to label regions, ``patches,'' of inputs as ``real'' (\textit{sampled} data) or ``fake'' (\textit{generated} data). Much of the research interest in pix2pix has centred on image translation tasks but the inherent structure of the U-Net model does not limit it to images alone. In fact original developments of U-Net were for semantic segmentation \cite{UNet}.

Research into GANs as they stand within the wider DGM and even wider generative ML/AI ecosystem have not been limited to images either. Parallel work on one-dimensional (1D) GANs where time series training data is periodic \cite{WaveGAN} has observed that derived models that decompose demonstrated two-dimensional models into 1D counterparts with a wider learning aperture, which we set ourselves with the size of convolution kernels, are capable of generating convincing high-accuracy 1D time series (including audio) from a learned signal data domain. Wider convolutional apertures are necessary for models to see and learn the time series periodicity. Others before have taken the conceptual essence of signal-to-signal translation and adapted its generator U-Net models for other signal domains; spectrum translation \cite{spec2spec} (spectral/frequency series-to-series), sensor translation \cite{sig2sig} (time series-to-series, 2D), and sound translation [\href{https://maxgraf.space/}{9}] (time series-to-series, 1D) to name a few. \cite{spec2spec} focused on the translation of sensor-to-sensor to adapt sensor conditions for object/material detection. \cite{sig2sig} focused on the re-use of sensor models, translating old sensor measurements to new ones; they used a modified 2D pix2pix-like architecture to do this. [\href{https://maxgraf.space/}{9}] focused on timbre-to-timbre translation in (optionally musical) audio signals, accomplishing this in 1D but again through what was actually a spectral-to-spectral translation like \cite{spec2spec}.

All of these share the idea of using machine learning to translate between types of windowed signals whether expressed in one or two-dimensional bases. The takeover of research interest in generic sequence-to-sequence and equivalently series-to-series translation models and their continued capacity to surpass many of the performance metrics that were previously maximized by more specialized convolutional neural networks and the like suggest that some model generality is possible. WaveGAN \cite{WaveGAN} showed that a general treatment of 2D convolutional models could make 1D convolutional counterparts suitable to generate audio with only a change of dimension and data shape. Still, where this generality starts and stops is unknown and to my knowledge little has been done to establish the generality of these models across 1D/2D signal paradigms.

\subsection{Objective.} This work aims to establish by existence proof that the practice of transcribing convolutional models between 2D to 1D and possibly other dimensional configurations has sufficient generality to solve several signal-to-signal translation problems. To this end I focus first on the 1D unpaired signal-to-signal translation task.

\section{Methods \& Materials}

\subsection{Model.}

In the interests of model re-usability I use a model that is readily available online: a modified CycleGAN \cite{CycleGAN} architecture that combines simplified versions of the generator and discriminator models (using only three U-Net down/upsampling layers) from the pix2pix \cite{pix2pix} architecture replacing 2D layers with 1D layers and widening convolutional kernels to roughly the square of their size. Notably CycleGAN uses the same generator and discriminator as pix2pix, differing in function only by training on paired/unpaired data and using a different training loss; CycleGAN combines two pix2pix models to learn signal translations to \textit{and} from both signal translation domains by minimizing ``roundtrip'' error.

This model was trained over the course of a few minutes using the standard CycleGAN losses, $\lambda = 10$, and $\beta_1 = 0.5$ for 100 epochs with a batch size of one using TensorFlow on an M1 MacBook Pro. Signals and their translated counterparts were scored using Pearson product-moment correlation ($r$-value) and mean absolute error (MAE).

\subsection{Dataset.}

A mock dataset simulates randomly windowed samples from a bandlimited signal where we can vary the window length, bandwidth, and maximum phase offset to study the capacities of the simplified 1D CycleGAN model.
I use a small collection of sixteen tunably (a)synchronous periodic bandlimited signals sampled reproducibly with a reusable randomness seed; each signal is made up of multiple sine waves with seed-generated random amplitudes, phase offsets (up to $2\pi$ radians), and frequency offsets. The (a)synchronicity of paired/unpaired signals is controllable through the extent of phase offsets. A sample of dataset elements is shown below.

\begin{figure}[h]
  \hypertarget{fig:1}{
    \centering
    \includegraphics[scale=0.5]{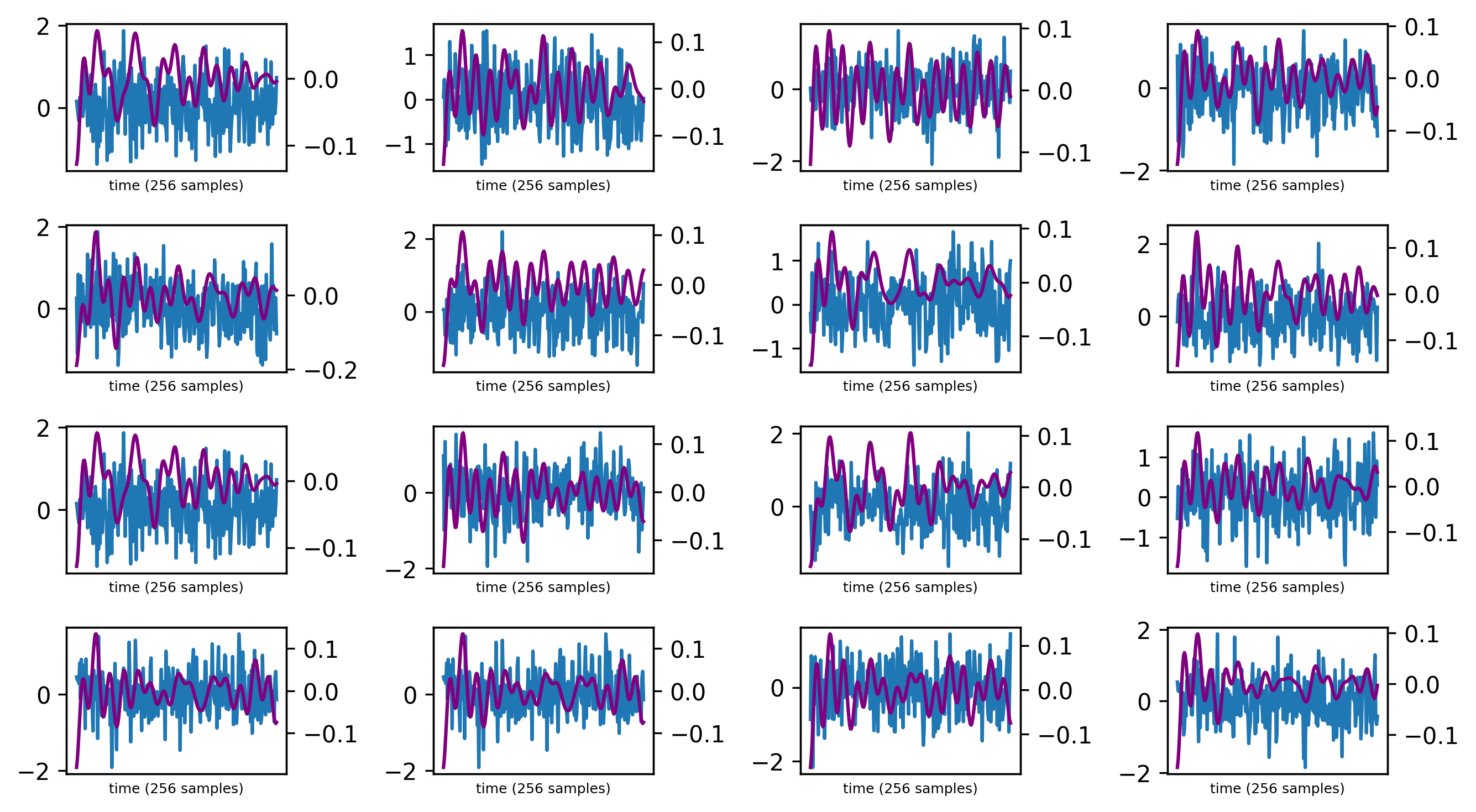}
  }
  \caption{Sixteen randomly sampled dataset elements used for 1D CycleGAN training between one-dimensional signal domains (blue/purple lines) with arbitrary scales.}
\end{figure}

\clearpage

\section{Results}
\begin{figure}[h]
  \hypertarget{fig:2}{
    \centering
    \includegraphics[scale=0.5]{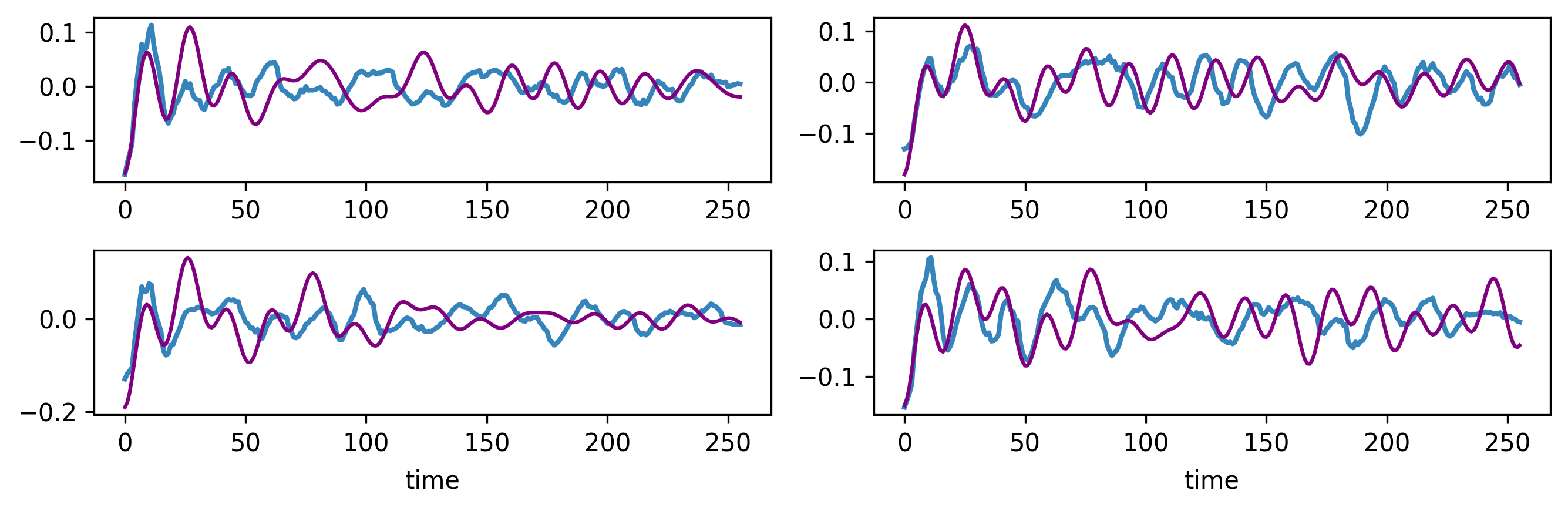}
  }
  \caption{Paired signal-to-signal translations by the simplified 1D CycleGAN architecture against a small unseen four-element test dataset in the \textit{time} domain ($r$-values in discussion).}
\end{figure}

\begin{figure}[h]
  \hypertarget{fig:3}{
    \centering
    \includegraphics[scale=0.5]{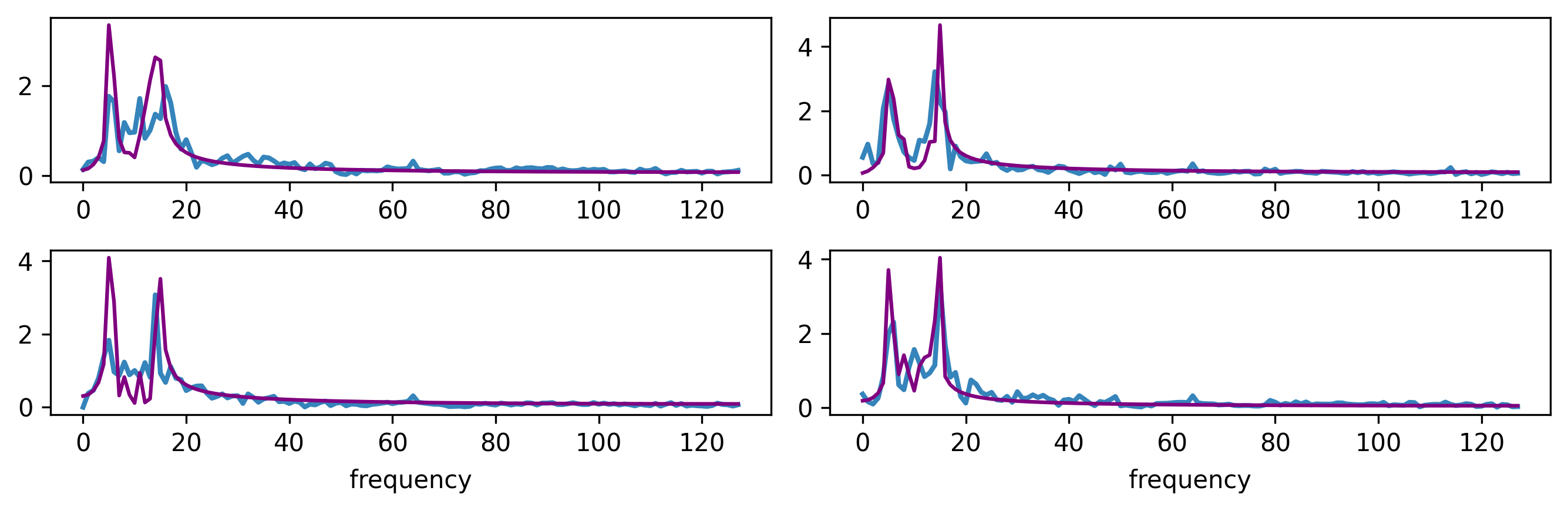}
  }
  \caption{Paired signal-to-signal translations by the simplified 1D CycleGAN architecture against a small unseen four-element test dataset in the \textit{frequency} domain after discrete Fast Fourier Transform. Frequency-wise signals seem to match more than time-wise signals above ($r$-values in discussion).}
\end{figure}

\begin{figure}[h]
  \hypertarget{fig:4}{
    \centering
    \includegraphics[scale=0.5]{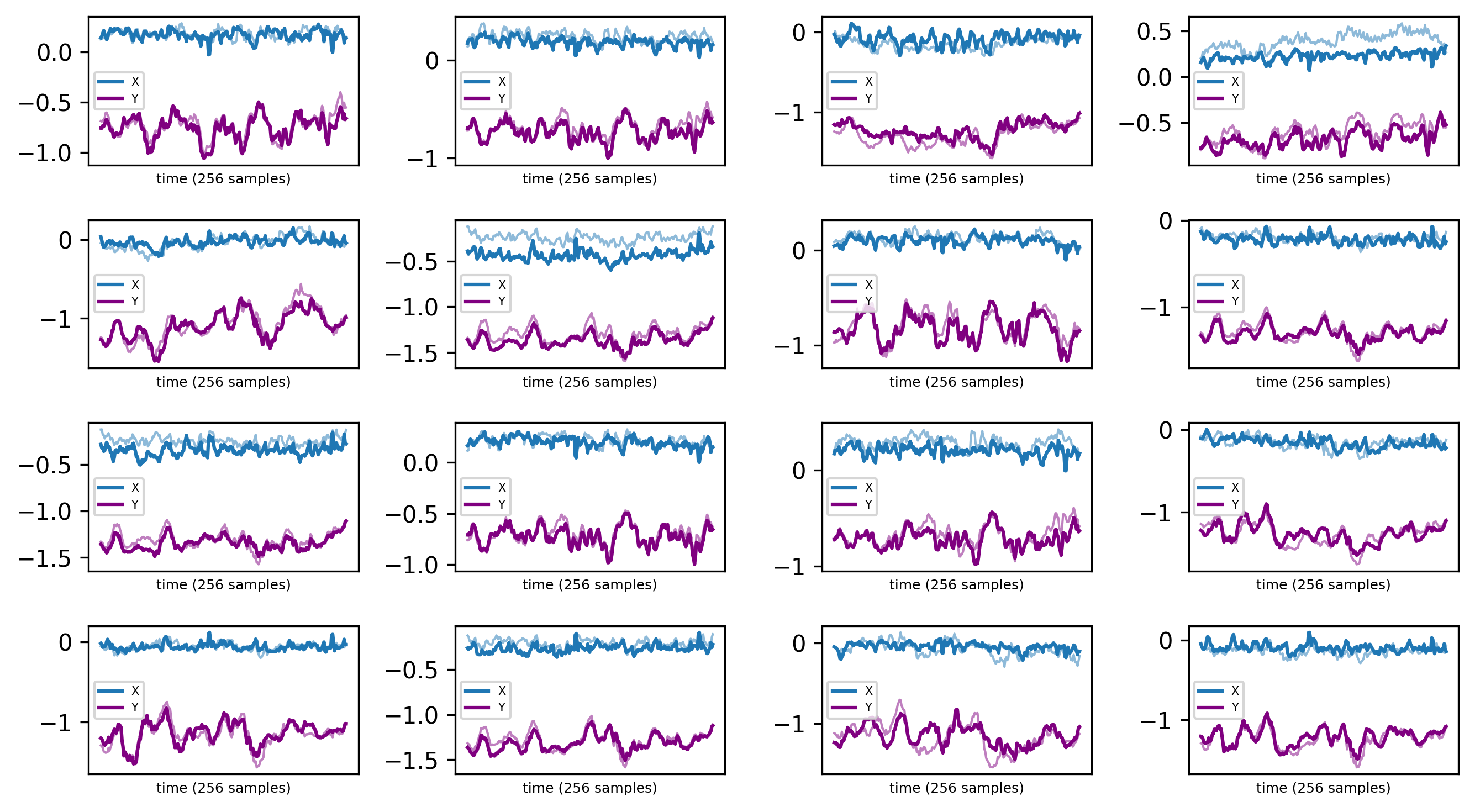}
  }
  \caption{Signal-to-signal translations (supplementary anonymized dataset) by the simplified CycleGAN architecture, which combines elements of the pix2pix architecture with a different training pattern and loss for unpaired signal data. Differences in reconstruction are shown at both signal $X$ and signal $Y$ translation domains (compared as solid/transparent lines).}
\end{figure}

\clearpage

\section{Discussion}

The simplified 1D CycleGAN model performs reasonably well with the synthetic dataset despite relatively little training (less than 10 minutes); for all four test samples (0.21 $<$ r $<$ 0.46 and (0.032 $<$ MAE $<$ 0.037) in the time domain and (0.71 $<$ r $<$ 0.89) and (0.13 $<$ MAE $<$ 0.15) in the frequency domain. Yet while the 1D CycleGAN model is trained on unpaired data it seems to be capable of accurately transforming between \textit{paired} signals never seen in training while also conserving much of their frequency content. This suggests interesting potential for CycleGAN and similar convolutional architectures to learn translations between (a)synchronous 1D signals.

Further testing with a larger synthetic data schema and several performance measures is needed to fully validate CycleGAN as a viable option for 1D signal-to-signal translation in more practical and wild data contexts.

\section{Conclusion}
I showed that a common generative architecture for 2D image-to-image translation model could be transformed into a 1D model by applying some of the same concepts as used by \cite{WaveGAN} to transform the 2D image DCGAN model \cite{DCGAN} into a 1D audio model. I showed that CycleGAN \cite{CycleGAN}, which applies the same principles as \cite{pix2pix} but with unpaired training data, performs with a small mock dataset of tunable periodic signals even without paired training suggesting some potential for CycleGAN and similar convolutional architectures to learn translations between (a)synchronous 1D signals.

\bibliographystyle{plain}
\bibliography{paper}

\begin{thebibliography}{1}

\bibitem{WaveGAN}
Chris Donahue, Julian McAuley, and Miller Puckette.
\newblock Adversarial {Audio} {Synthesis}, February 2019.
\newblock arXiv:1802.04208 [cs].

\bibitem{GAN}
Ian~J. Goodfellow, Jean Pouget-Abadie, Mehdi Mirza, Bing Xu, David Warde-Farley, Sherjil Ozair, Aaron Courville, and Yoshua Bengio.
\newblock Generative {Adversarial} {Networks}, June 2014.
\newblock arXiv:1406.2661 [cs, stat].

\bibitem{pix2pix}
Phillip Isola, Jun-Yan Zhu, Tinghui Zhou, and Alexei~A. Efros.
\newblock Image-to-{Image} {Translation} with {Conditional} {Adversarial} {Networks}, November 2018.
\newblock arXiv:1611.07004 [cs].

\bibitem{sig2sig}
SangYeon Kim, Hyunwoo Lee, Jonghee Han, and Joon-Ho Kim.
\newblock {Sig2Sig}: {Signal} {Translation} {Networks} to {Take} the {Remains} of the {Past}.
\newblock In {\em {ICASSP} 2021 - 2021 {IEEE} {International} {Conference} on {Acoustics}, {Speech} and {Signal} {Processing} ({ICASSP})}, pages 3620--3624, June 2021.
\newblock ISSN: 2379-190X.

\bibitem{UNet}
Jonathan Long, Evan Shelhamer, and Trevor Darrell.
\newblock Fully {Convolutional} {Networks} for {Semantic} {Segmentation}, March 2015.
\newblock arXiv:1411.4038 [cs].

\bibitem{spec2spec}
Cara~P. Murphy and John Kerekes.
\newblock {1D} conditional generative adversarial network for spectrum-to-spectrum translation of simulated chemical reflectance signatures.
\newblock {\em Journal of Spectral Imaging}, 10, June 2021.

\bibitem{DCGAN}
Alec Radford, Luke Metz, and Soumith Chintala.
\newblock Unsupervised {Representation} {Learning} with {Deep} {Convolutional} {Generative} {Adversarial} {Networks}, January 2016.
\newblock arXiv:1511.06434 [cs].

\bibitem{CycleGAN}
Jun-Yan Zhu, Taesung Park, Phillip Isola, and Alexei~A. Efros.
\newblock Unpaired {Image}-to-{Image} {Translation} using {Cycle}-{Consistent} {Adversarial} {Networks}, August 2020.
\newblock arXiv:1703.10593 [cs].

\end{thebibliography}

\end{document}